%% file: main.tex
\documentclass[runningheads]{llncs}

\usepackage{graphicx}
\usepackage{hyperref}
\usepackage{booktabs}
\usepackage{multirow}
\usepackage{paralist}
\usepackage{xcolor,colortbl}
\usepackage{cite}
\usepackage{amsmath}
\usepackage{nicefrac}
\usepackage{todonotes}
\usepackage{xspace}

\definecolor{Gray}{gray}{0.9}
\newcolumntype{g}{>{\columncolor{Gray}}r}
\newcolumntype{h}{>{\columncolor{Gray}}c}
\newcommand{\pp}[1]{\vspace{6pt}\noindent\textbf{\emph{#1 --}}\xspace}

\begin{document}

\title{Unveiling User Behavior on Summit\\Login Nodes as a User}


\author{
    Sean R.\ Wilkinson\orcidID{0000-0002-1443-7479} \and
    Ketan Maheshwari\orcidID{0000-0001-6852-5653} \and
    Rafael Ferreira da Silva\orcidID{0000-0002-1720-0928}
}
\authorrunning{S.\ R.\ Wilkinson \and K.\ Maheshwari \and R.\ Ferreira da Silva}

\institute{
    Oak Ridge National Laboratory, Oak Ridge, TN, USA\footnote{\scriptsize This manuscript has been authored by UT-Battelle, LLC, under contract DE-AC05-00OR22725 with the US Department of Energy (DOE). The publisher acknowledges the US government license to provide public access under the DOE Public Access Plan (http://energy.gov/downloads/doe-public-access-plan).}
    \\
    \email{
        \{wilkinsonsr,maheshwarikc,silvarf\}@ornl.gov
    }
}

\maketitle

\begin{abstract}
\input{abstract}


\end{abstract}

\input{sec-introduction}
\input{sec-characteristics}
\input{sec-metrics}

\input{sec-userbehavior}

\input{sec-relatedwork}
\input{sec-conclusion}

{\medskip
\small
\noindent \textbf{\emph{Acknowledgments.}}
This research used resources of the Oak Ridge Leadership Computing Facility at
the Oak Ridge National Laboratory, which is supported by the Office of Science
of the U.S. Department of Energy under Contract No. DE-AC05-00OR22725. We
acknowledge Suzanne Parete-Koon for early brainstorming of some of the ideas
presented here. We thank Scott Atchley, Bronson Messer, and Sarp Oral for their
thorough revision of this paper.
}


\end{document}

%% file: abstract.tex


We observe and analyze usage of the login nodes of the leadership class Summit
supercomputer from the perspective of an ordinary user---not a system
administrator---by periodically sampling user activities (job queues, running
processes, etc.) for two full years (2020--2021). Our findings unveil key usage
patterns that evidence misuse of the system, including gaming the policies,
impairing I/O performance, and using login nodes as a sole
computing resource. Our analysis highlights observed patterns for the execution
of complex computations (workflows), which are key for processing large-scale
applications.


%% file: sec-introduction.tex
\section{Introduction}

HPC systems have been designed to address computing, storage, and networking
needs for complex, high-profile applications.  Specifically,
leadership class supercomputers~\cite{top500} meet the needs of applications
that require high-speed interconnects, low latency, high I/O throughput, and
fast processing capabilities (currently petascale, and soon
exascale)~\cite{dongarra2019race}. Understanding the performance of these
systems and applications is a cornerstone for the design and development of
efficient, reliable, and scalable systems. To this end, several works have
focused on the development of system- and application-level monitoring and
profiling tools that can provide fine-grained
characterizations of systems' and applications' performance.

The current landscape of HPC systems performance research is mostly focused on
the system's performance---which is utterly valuable for systems
design~\cite{ferreiradasilva-iccs-2015, liu2020characterization, bang2020hpc}.
However, the \emph{user perception} of the system is often disregarded, and
there is a common misconception that application execution performance is the
only consideration for user satisfaction. Although application performance is
one of the chief goals of HPC, there are several additional factors that impact
user experience. More specifically, before experiencing the capabilities of the
HPC nodes, users' first interactions are with the \emph{login nodes},
where users share resources like CPU, memory, storage, and network bandwidth
while performing basic tasks like compiling code, designing experiments, and
orchestrating services. The login nodes on an HPC system represent a gateway to
the system which is often overlooked when considering the capabilities and
performance of the overall system. We argue that the experience on the login
nodes may impact a user's perception of and behavior on the system, thus
influencing whether and how the user continues to utilize that system in future
work.

\begin{figure}[!t]
    \centering
    \vspace{-15pt}
    \includegraphics[width=\linewidth]{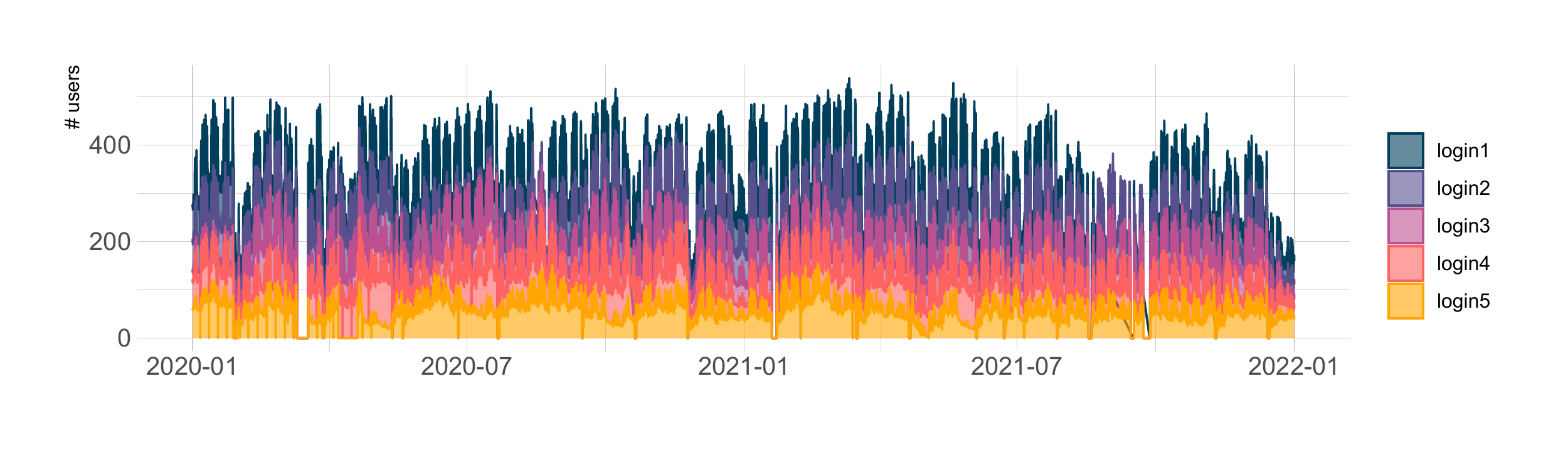}
    \vspace{-30pt}
    \caption{Distribution of users over 5 Summit login nodes
        (Jan 2020--Dec 2021). Gaps in the time series indicate outages or
        system downtime.}
    \label{fig:users-dist}
\end{figure}

In this paper, we attempt to identify long-term usage patterns by collecting 
observational data on the login nodes from the Summit leadership class 
supercomputer hosted at the Oak Ridge Leadership Computing Facility (OLCF) at
Oak Ridge National Laboratory (ORNL). Every hour for two years (2020--2021), we
have collected data about the login nodes' performance with respect to CPU,
memory, and disk usage, and we also collected data about their activity with
respect to logged-in users, the programs they were running on the login nodes,
and the status of all user jobs. Figure~\ref{fig:users-dist} shows the number
of users per login node within this time window. In addition to examining
traditional metrics like system usage and distribution of jobs and users across
the nodes, we also seek to (i)~highlight atypical usage and user misconducts,
(ii)~relate these behaviors to potential performance issues, (iii)~identify
usage patterns of a complex class of applications such as scientific workflows,
and (iv)~establish relationships between users' sessions length and system
load.

%% file: sec-characteristics.tex
\section{Characteristics of the Summit Login Node Data}
\label{sec:characteristics}

\begin{figure}[!t]
    \centering
    \includegraphics[width=\linewidth]{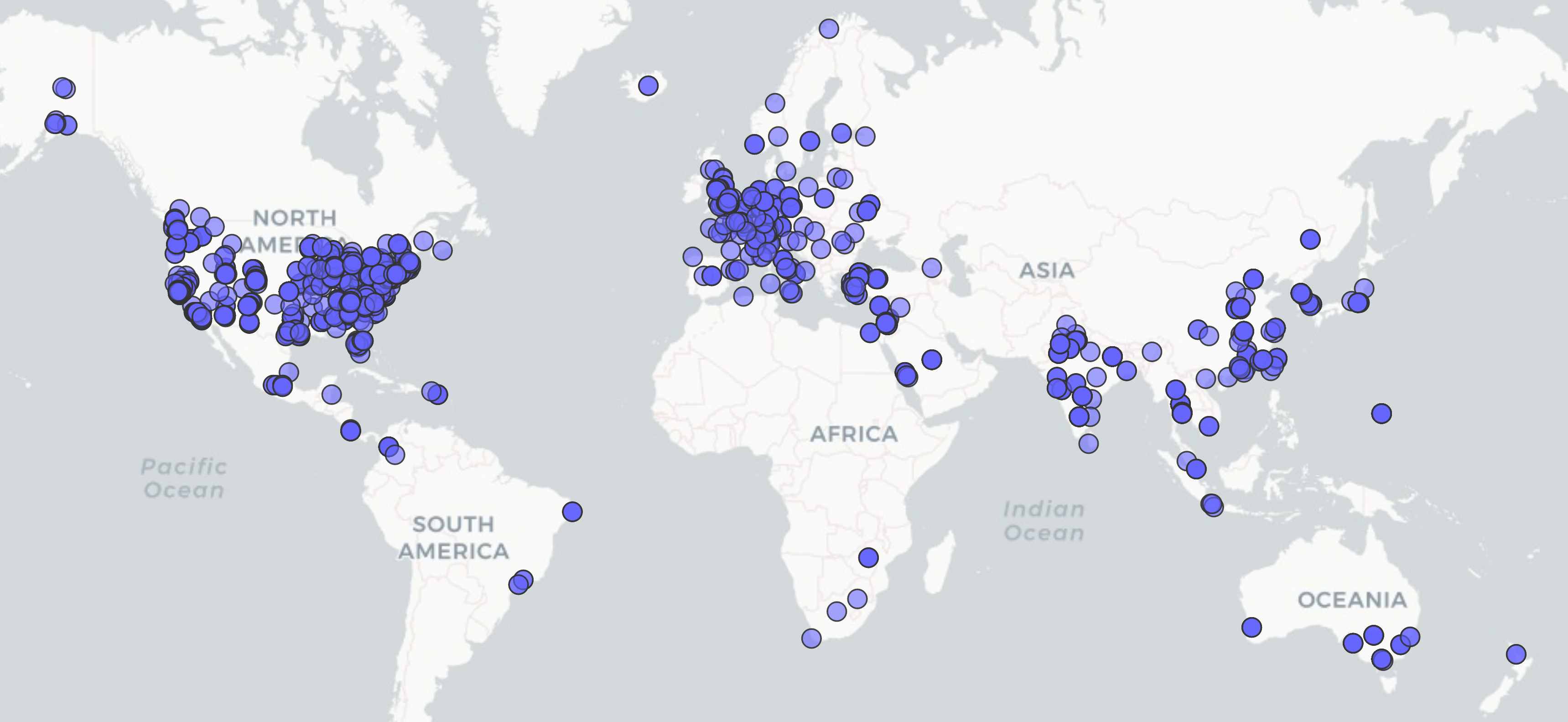}
    \vspace{-20pt}
    \caption{Users' geolocation distribution obtained with IP lookup
        ($\sim$93\% of total users).}
    \label{fig:map}
\end{figure}

Table~\ref{tab:summary} summarizes the main characteristics of the collected 
data. The dataset represents activity from 1,967 unique users, who connected
using 9,841 unique IPs and submitted 1,783,867 jobs, of which 1,073,754
completed successfully while 705,103 had a non-zero exit code. 
Figure~\ref{fig:map} shows the distribution of users' geolocations, which
were resolved through an IP geolocation tool~\cite{abstractapi}. For the sake
of privacy, any user-specific data had been previously anonymized and not
retained.

\begin{table}[!ht]
    \centering
    \setlength{\tabcolsep}{6pt}
    \scriptsize
    \caption{Characteristics of Summit login node data for a period 
             of two years (Jan 2020--Dec 2021). Totals for ``Unique Users'' and
            ``Unique IPs'' do not sum additively due to Summit users whose use
            spanned both years. Additionally, the total number of jobs may not
            coincide with the sum of individual years because jobs may be 
            carried over from one year to the next.}
    \begin{tabular}{lrrrrrr}
        \toprule
        \multirow{2}{*}{Year} & \multirow{2}{*}{\# Unique Users} & \multirow{2}{*}{\# Unique IPs} & \multicolumn{4}{c}{\# Jobs} \\
        & & & completed & suspended & exited & total \\
        \midrule
        2020 & 1,509 & 5,094 & 480,550 & 1,869 & 313,257 & 795,676 \\        
        2021 & 1,514 & 5,467 & 668,580 & 3,264 & 410,493 & 1,082,337 \\
        \midrule
        Total & 1,967 & 9,841 & 1,073,754 & 5,010 & 705,103 & 1,783,867 \\
        \bottomrule
    \end{tabular}
    \label{tab:summary}
\end{table}

\pp{System Characteristics and Data Collection}
Summit is equipped with 5 login nodes~\cite{vazhkudai2018design}. Each login
node runs Red Hat Enterprise Linux v8.2 and comes equipped with two 3.8 GHz
16-core IBM POWER9 CPUs (4 threads per core), 512 GiB of DDR4-2667, 4 NVidia
V100 GPUs each with 16 GiB of HBM2, and connection to a 250 PB GPFS scratch
filesystem. Users usually log into Summit via SSH to the load-balanced
\texttt{summit.olcf.ornl.gov} hostname, but they can optionally connect to a
specific login node. Data were collected hourly, starting January 1, 2020, on
all five login nodes. A shell script ran in the user space as a \texttt{while}
loop within a Linux \texttt{tmux} session because user cronjobs are not
allowed, and it collected traditional system usage performance metrics as well
as user behavior (e.g., running processes and jobs). One caveat is that the
hourly sampling frequency may have failed to capture fine grained behavior, as
many things can happen between samples. Nevertheless, we believe that the large
volume of samples sufficiently captures most of the representative system and
user activity. More precisely, each sample collects the following data:

\begin{compactitem}
    \item List of currently logged-in users using the \texttt{w} command;
    \item CPU and memory usage using the \texttt{top} and \texttt{ps} 
          commands (which also provides the list of running processes), and 
          statistics from \texttt{meminfo} and \texttt{vmstat} in the
          \texttt{/proc} filesystem;
    \item Status of users' batch jobs via the \texttt{bjobs} command;
    \item Disk usage statistics using the \texttt{df -h} command and disk
          throughput by measuring the timespan for writing a 1GB data file to
          GPFS.
\end{compactitem}

\pp{Data Preparation}
Real-world data may be incomplete, noisy, and inconsistent, which can obscure 
useful patterns~\cite{zhang2003data}. Data preparation techniques cannot be 
fully automated; it is necessary to apply them with knowledge of their effect
on the data being prepared. We used our prior knowledge about the execution of
scientific applications on HPC to extract and combine relevant information from
each source of data. We have then pre-processed the dataset by removing
redundancies and missing data (e.g., due to outages and system downtimes),
sanitizing lists of programs and users for long-running processes and jobs, and
resolving IP addresses for filtering and identifying individual users and their
locations, among other things.

%% file: sec-metrics.tex
\section{System Metrics}
\label{sec:metrics}

In this section, we examine overall characteristics and performance metrics
from Summit. The assessed set of metrics are restricted to an ordinary user's
perspective of the system, as viewed from a login node. Although these metrics
are often reported and analyzed in-depth from the system's perspective by using
system-wide monitoring and profiling tools, here we have used a subset of these
metrics to support our claims regarding user experience and behavior. 

\subsection{Users Access}
\label{sec:access}

\begin{figure}[!t]
    \centering
    \vspace{-15pt}
    \includegraphics[width=\linewidth]{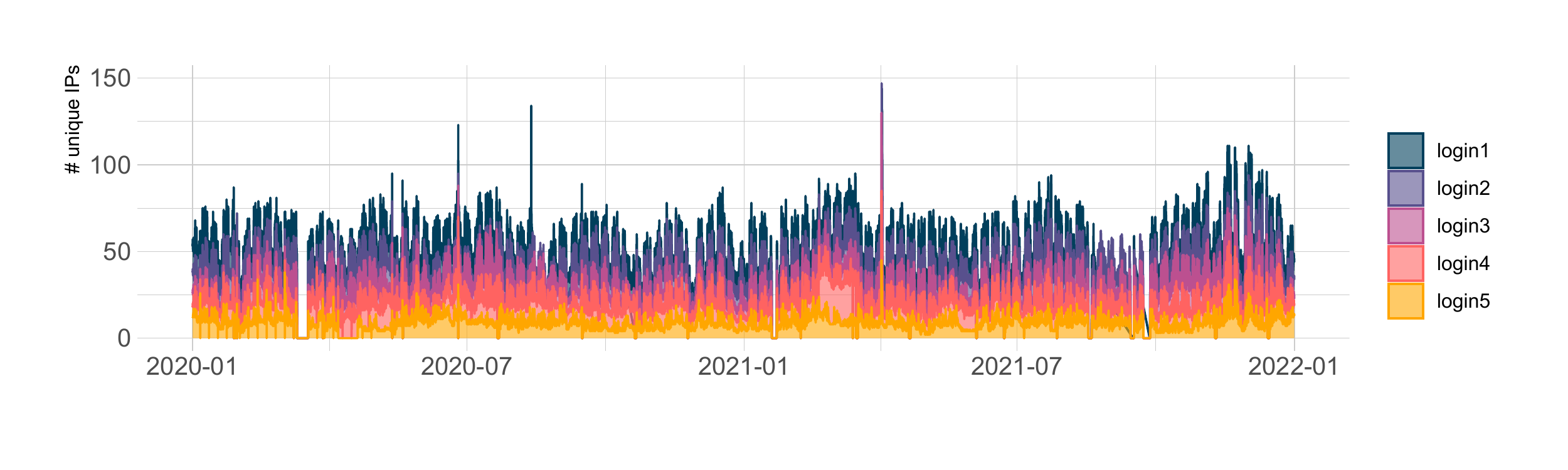}
    \vspace{-30pt}
    \caption{Distribution of unique IPs across Summit login nodes
        (Jan 2020--Dec 2021).}
    \label{fig:unique-ips}
\end{figure}

Figure~\ref{fig:users-dist} shows the distribution of user sessions per login 
node. The average percentage of user distribution is 21.9\% ($\pm$9.2\%),
20.7\% ($\pm9.2\%$), 17.9\% ($\pm$7.8\%), 20.2\% ($\pm$8.8\%), and 19.3\% 
($\pm$8.1\%) for login nodes 1--5, respectively. Although this distribution 
is relatively balanced among login nodes, by inspecting the distribution of 
unique IPs per login session (Figure~\ref{fig:unique-ips}) we observe that 
there is an imbalance on the disposition of individual users among the nodes.
Specifically, the average percentage of unique IPs distribution is 20.5\% 
($\pm$8.8\%), 20.8\% ($\pm$9.2\%), 16.7\% ($\pm$8.6\%), 23.5\% ($\pm$10.1\%),
and 18.5\% ($\pm$8.1\%) for login nodes 1--5, respectively. This 
indicates that a subset of users may be (involuntarily) benefitting from an 
increased number of concurrent login sessions; thus, their perceived experience 
of the system may be more favorable when compared with users who share
resources with a larger number of individual users.

\begin{figure}[!t]
    \centering
    \vspace{-15pt}
    \includegraphics[width=\linewidth]{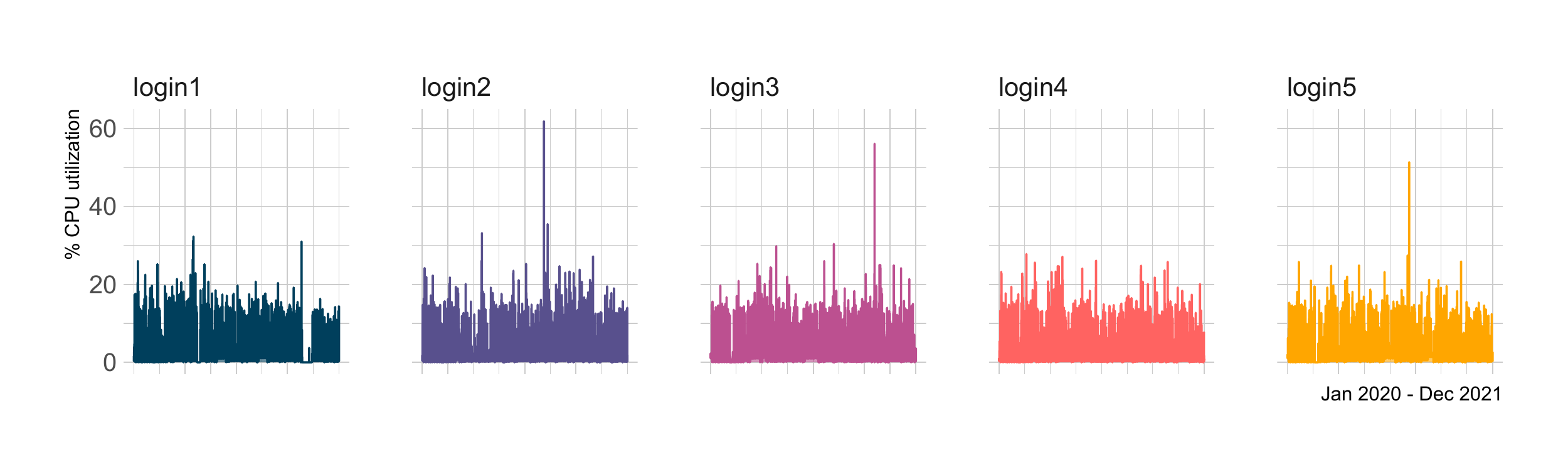}
    \vspace{-30pt}
    \caption{CPU utilization on Summit login nodes (Jan 2020--Dec 2021).}
    \label{fig:cpu}
\end{figure}

\begin{figure}[!t]
    \centering
    \vspace{-15pt}
    \includegraphics[width=\linewidth]{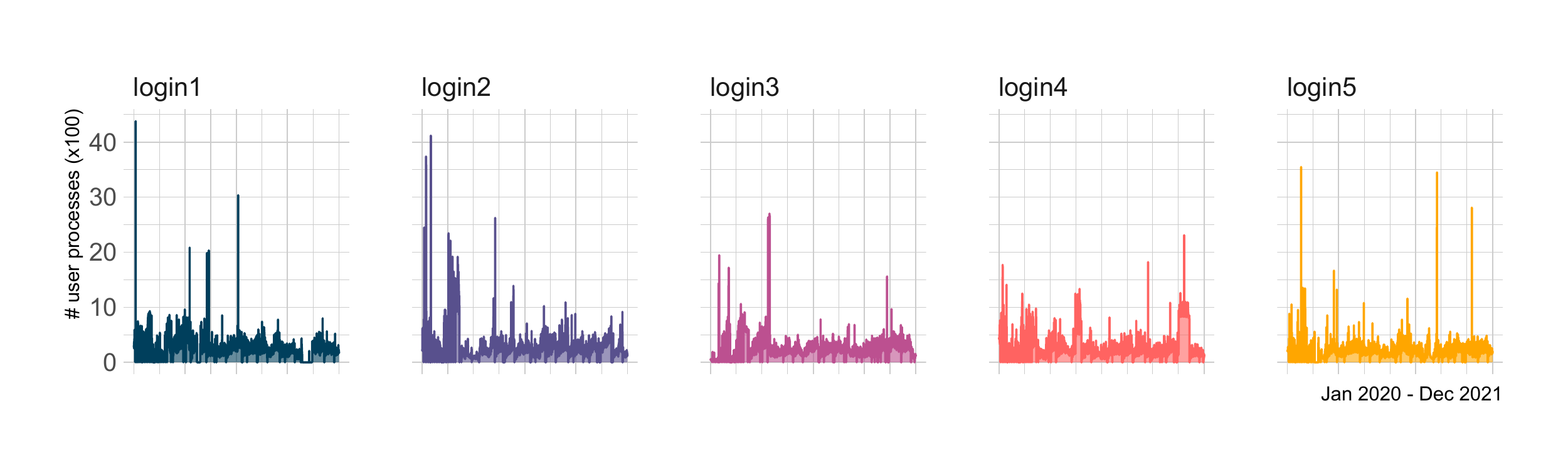}
    \vspace{-30pt}
    \caption{User processes on Summit login nodes (Jan 2020--Dec 2021).}
    \label{fig:uprocs}
\end{figure}

To evaluate the above claim, we examined CPU utilization and the
number of user processes per login node (Figures~\ref{fig:cpu}
and~\ref{fig:uprocs}). Overall, CPU utilization is relatively balanced among 
nodes (around 15\% in average across nodes) with some spikes on login nodes 
2, 3, and 5. Unsurprisingly, the number of user processes follows similar
trends as for the distribution of unique IPs. Both of these results corroborate
the claim that a small subset of users have been benefited from lower
concurrency. More precisely, the balanced distribution of CPU utilization on
login nodes 3 and~5 indicates that this small set of users consumes as many
resources on these nodes as the larger set of users on the other nodes.

\subsection{I/O Throughput}

\begin{figure}[!t]
    \centering
    \vspace{-15pt}
    \includegraphics[width=\linewidth]{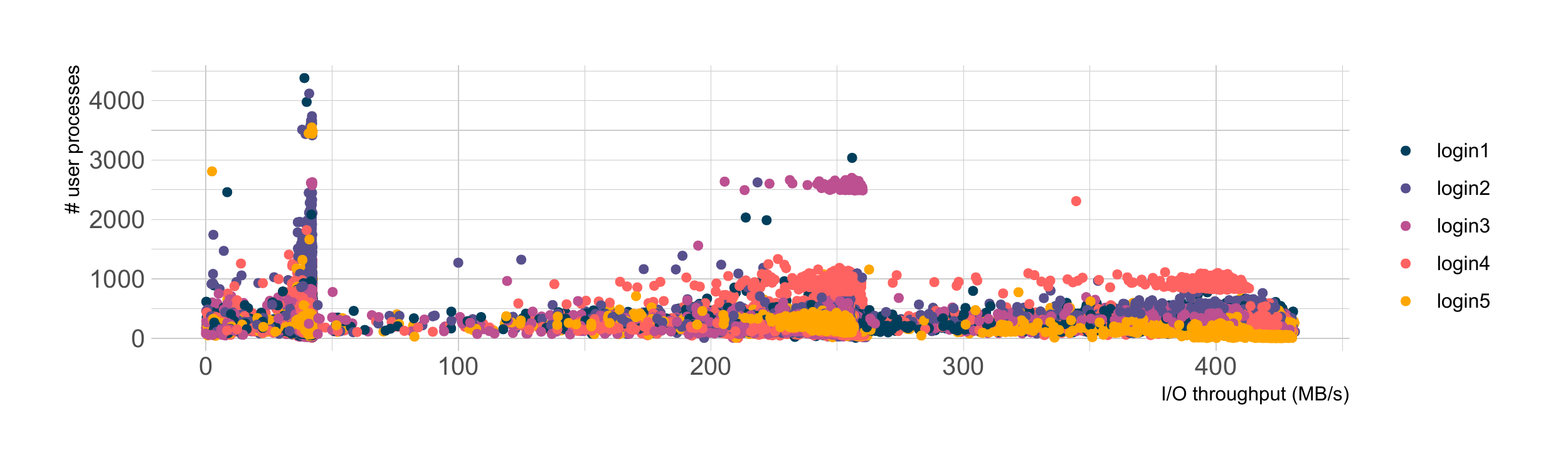}
    \vspace{-30pt}
    \caption{I/O throughput of Summit's GPFS for writing a 1 GB file in
        relation to the number of user processes.}
    \label{fig:io}
\end{figure}

Every hour, we have measured the I/O throughput of Summit's GPFS for writing a
1 GB randomly generated binary data file to a shared folder. Notice that we do
not aim to assess peak write speeds; instead our goal is to identify potential
low performance caused by user-related I/O operations within the login nodes.
Figure~\ref{fig:io} shows the distribution of the number of user processes
running per login node in relation to the I/O throughput for writing a 1 GB
file. Note that the performance of the GPFS filesystem may also be affected by
I/O operations occurring on the compute nodes; thus a weak correlation is
expected with processes running on the login nodes. That said, we can observe
that a low performance is highly correlated with an increased number of user
processes running on the login nodes. Specifically, throughput values as low as
42MB/s are reported when more than 3,000 user processes are running for more
than 3 consecutive hours. For the same set of datapoints, user processes
running on login5 run for more than 5 consecutive hours, which coincides with
the timespan in which the filesystem yields low performance (recall that login5
has, on average, a reduced number of concurrent unique users). Analogously,
impaired performance (around 250 MB/s) is observed for a very small group of
users who run more than 2,500 processes on login3 for more than 7 hours.

\subsection{Computational Jobs}
\label{sec:jobs}

\begin{figure}[!t]
    \centering
    \vspace{-10pt}
    \includegraphics[width=\linewidth]{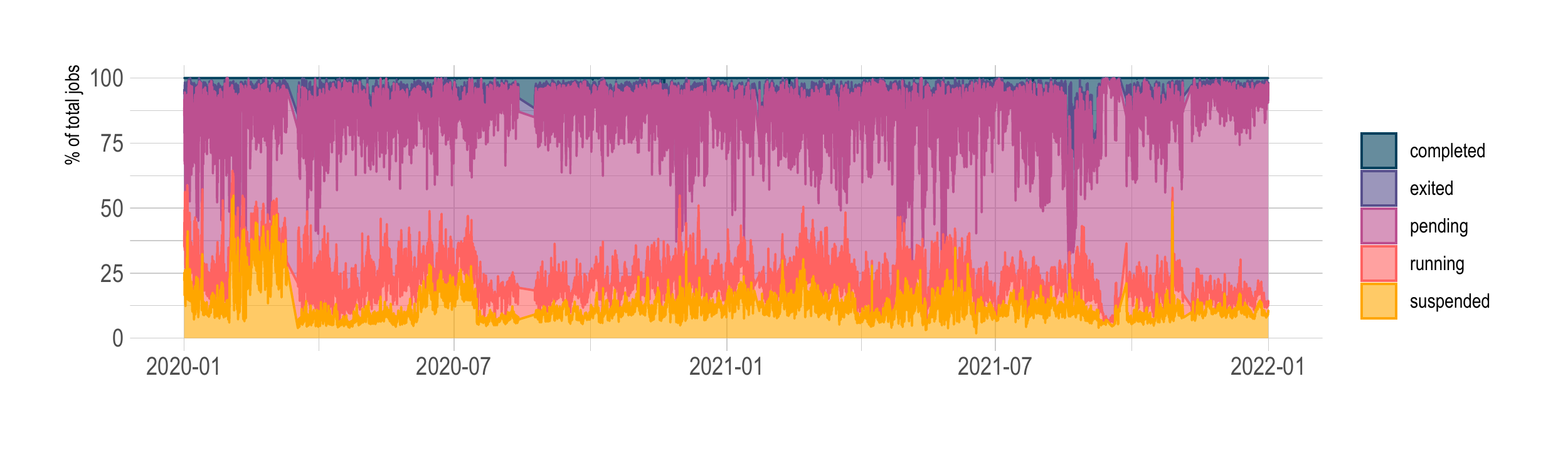} \\ \vspace{-25pt}
    \includegraphics[width=\linewidth]{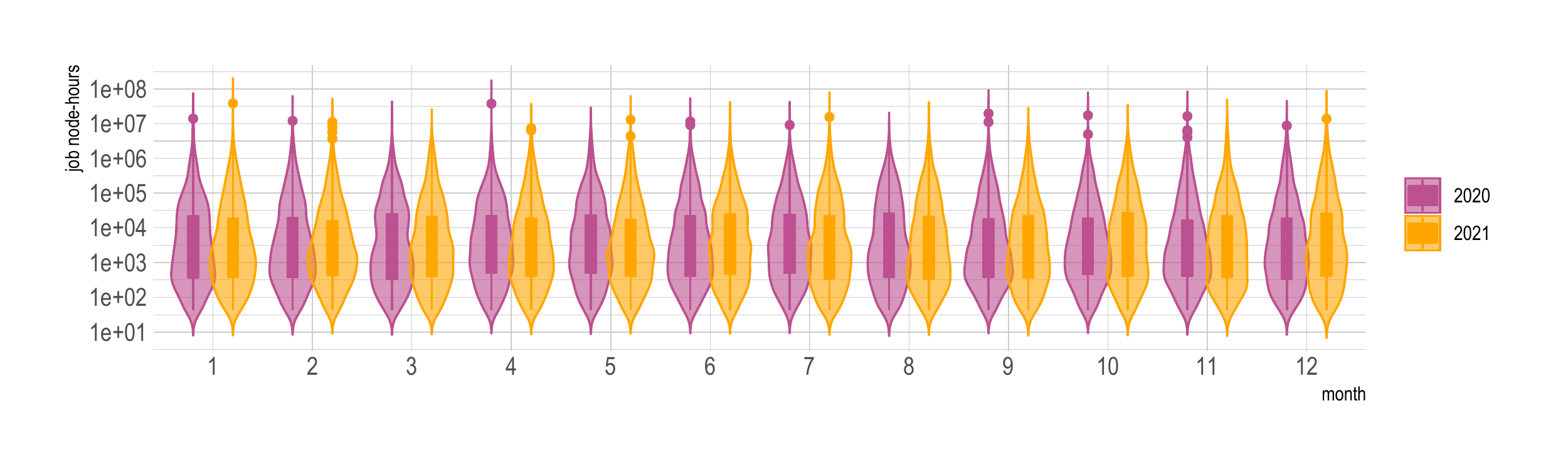}
    \vspace{-30pt}
    \caption{Distribution of jobs' statuses (\emph{top}) and jobs' sizes
        (\emph{bottom}). Each ``violin'' represents the distribution of jobs'
        sizes in a given month in terms of node-hours as a rotated kernel
        density plot on top of a box-plot that shows the first and third
        quartiles of the distribution; the width of the violin corresponds to
        the number of jobs, and the dots indicate outliers in the tails.}
    \label{fig:jobs}
\end{figure}

The fundamental purpose of leadership-class supercomputers is to improve
science by running the largest-scale computational jobs. It is expected that
user satisfaction is mostly dictated by the ability to execute batch jobs
successfully with good performance and without long waits in the queue.
Figure~\ref{fig:jobs}-\emph{top} shows the percentage distribution of jobs
based on their status. The workload average jobs submitted, running, and
completed, as shown by the LSF scheduler, are 529 ($\pm$271), 81 ($\pm$24), and
90 ($\pm$69), respectively. Given that the number of individual users
(see Table~\ref{tab:summary}) is orders of magnitude higher than the average
number of running jobs, the variation in the number of running jobs seems
relatively low.

\begin{figure}[!t]
    \centering
    \vspace{-10pt}
    \includegraphics[width=\linewidth]{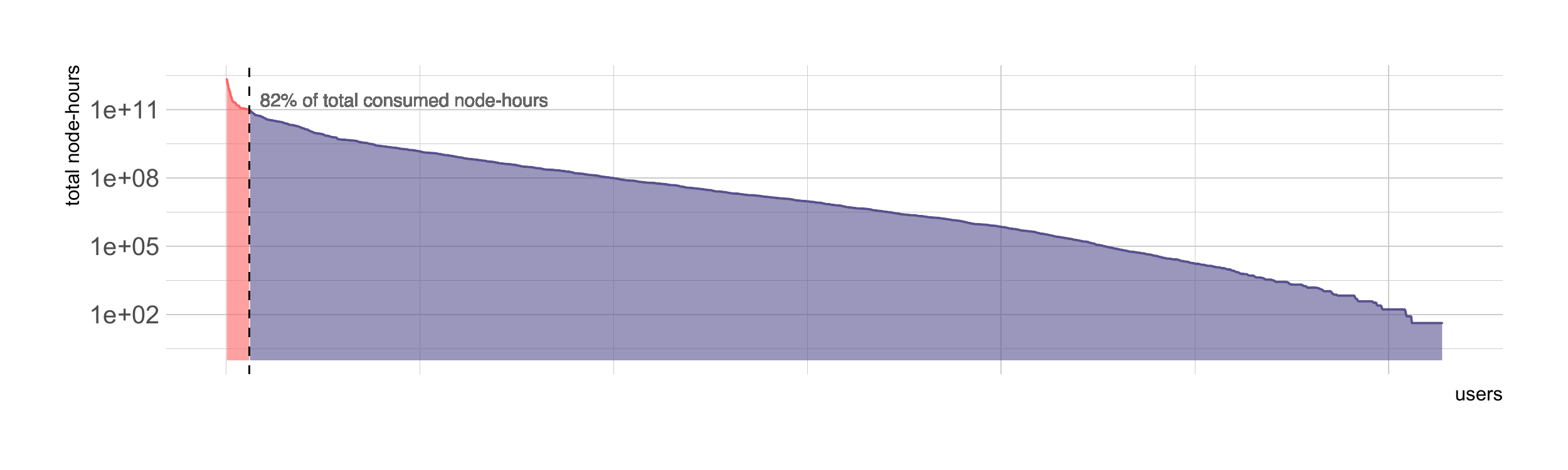}
    \vspace{-20pt}
    \caption{Distribution of total node-hours consumed per user. (The dashed
             vertical line delineates the total node-hours consumed by 29 
             users, which represents more than 82\% of the entire dataset.)}
    \label{fig:nodehours}
\end{figure}

Figure~\ref{fig:jobs}-\emph{bottom} shows the distribution of node-hours 
consumed per job. Intriguingly, the shape of the distributions are alike
across years and months. More precisely, the average root mean square error 
(RMSE) is below 6 for every month comparison between the two years, with most 
jobs consuming between 1,000 and 10,000 node-hours. This result suggests that 
jobs are mostly submitted by a small set of users running similar, yet large, 
workloads. Indeed, by examining the number of jobs submitted per user,
we observe that \emph{only 29 users ($\sim$1.4\% of total number of users in
the dataset), submitted more than 50\% of all Summit jobs over the measured
period of time.} These jobs represent more than 82\% of the total consumed
node-hours in the dataset (Figure~\ref{fig:nodehours}). As expected, most
individual jobs consume between 1,000 and 10,000 node-hours, which corroborates
the findings asserted from Figure~\ref{fig:jobs}-\emph{bottom}. Most users
submitted a very small number of jobs, though they span a wide range of
node-hours consumption, with a few jobs consuming nearly all available compute 
resources. We can also observe that specific users submitted sets of individual 
jobs with a wide range of node-hours (e.g., from 96 up to 193,537), but also 
submitted more than 1,100 jobs with the same size (e.g., $\sim$4,300 node-hours).

%% file: sec-userbehavior.tex
\section{User Behavior}
\label{sec:userbehavior}

HPC performance metrics are traditionally associated with success metrics such
as high system utilization and large number of users and jobs, which correlate
to wide system adoption by the community and fulfillment of scientific goals.
Understanding and modeling user behavior in HPC environments is key to
exhibiting usage patterns that may help improve the design of the system,
relate performance bottlenecks to specific behaviors, and ascertain violations
of policies and best practices, among other things. Previous studies have
mainly focused on job characteristics (performance metrics as presented in
Section~\ref{sec:jobs}) and scheduling (queuing time, wall time,
etc.)~\cite{schlagkamp-iccs-2016, rodrigo2018towards, wolter2006s}. In this
section, we examine user behavior from the standpoint of (i)~the average user
session length, (ii)~misuses, and (iii)~usage patterns of a complex class of
applications such as scientific workflows.

\subsection{User Sessions}

In this section, we investigate the length of user sessions in an attempt to 
characterize user behavior by relating the time users spend logged into the
system with the number and size (in terms of number of nodes) of jobs
submitted. We define a \emph{session} as a time interval indicated by activity
which begins and ends with inactivity. We use batch job submission as the
indicator of activity, and for inactivity, we leverage
\emph{think time}~\cite{feitelson2008looking}, which quantifies the time
between the completion of a job and the submission of the next job by the same
user. Thus, a session is the time period that complements two subsequent think
times for the same user. In this work, we assume that a think time is
characterized by an interval of more than 24 hours. We do not consider
weekends, holidays, or system downtimes or outages as think times.

\begin{figure}[!t]
    \centering
    \vspace{-10pt}
    \includegraphics[width=\linewidth]{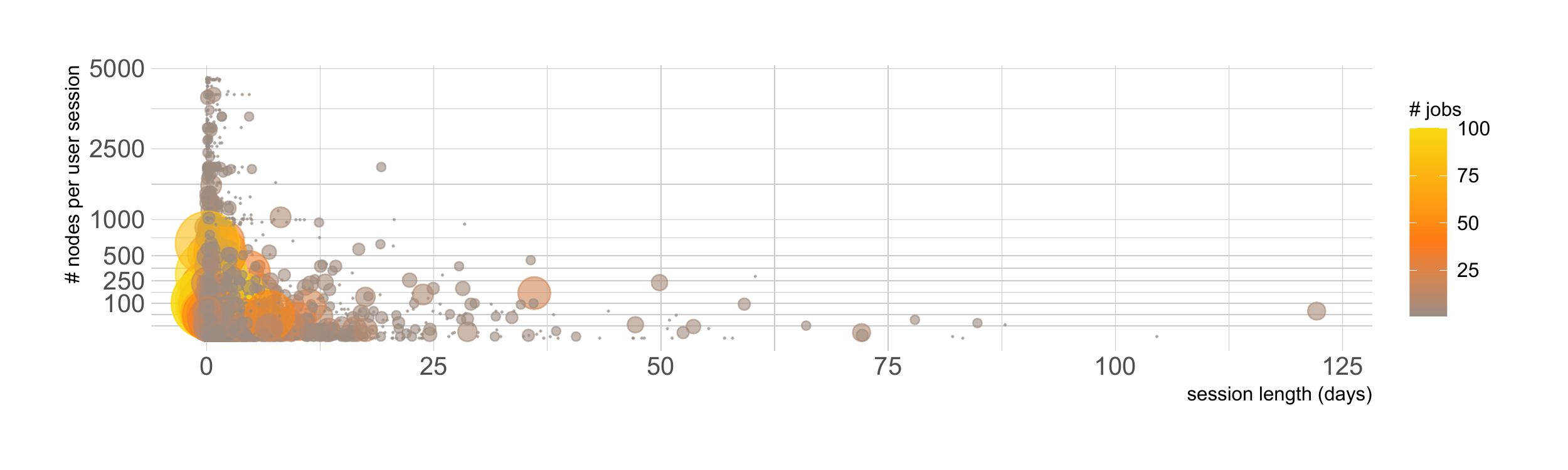}
    \vspace{-25pt}
    \caption{Users' session lengths (in days) in relation to the number of jobs
        submitted and maximum number of nodes requested.}
    \label{fig:sessions}
\end{figure}

We identified 27,789 sessions, the longest of which spans 123 days and runs 68
jobs over a maximum of 64 nodes. Most of the users (about 92\%) established
more than one session, and most of the user sessions (about 84\%) span less
than one day; also, more than 50\% of these sessions request only 1 or 2 nodes
per job. Sessions with large-scale jobs that use nearly all of Summit's compute
nodes span only a few hours, with only 3 spanning slightly more than one day.
This supports the idea that user experience on login nodes significantly
impacts user satisfaction, because users spend most of their time testing and
debugging while using the login nodes. Figure~\ref{fig:sessions} shows the
distribution of user sessions' lengths in relation to the total number of nodes
used by all jobs within a session.

\subsection{Misuse}

Typically, HPC systems balance users across the set of login nodes to improve
the overall user experience and limit any potential performance impact due to
heavy user processes (see Section~\ref{sec:access}). To prevent low quality of
service, most HPC systems provide guidance and best practices for operations
that should not be performed on login nodes because they are shared resources.
For instance, it is discouraged to run long-term and/or heavy services (e.g.,
databases) on such nodes. In this section, we examine whether users run
processes that could harm the overall performance of these shared resources. To
this end, we mined the dataset for processes that did not represent typical,
system-related tasks, that consumed a substantial amount of resources (CPU /
GPU / memory), or that ran for a long period of time. We limit our discussion
in this section to two representative use cases: (i)~execution of tightly
coupled applications using \texttt{mpirun} and \texttt{mpiexec}, and
(ii)~execution of high-throughput applications.

\begin{figure}[!t]
    \centering
    \vspace{-10pt}
    \includegraphics[width=\linewidth]{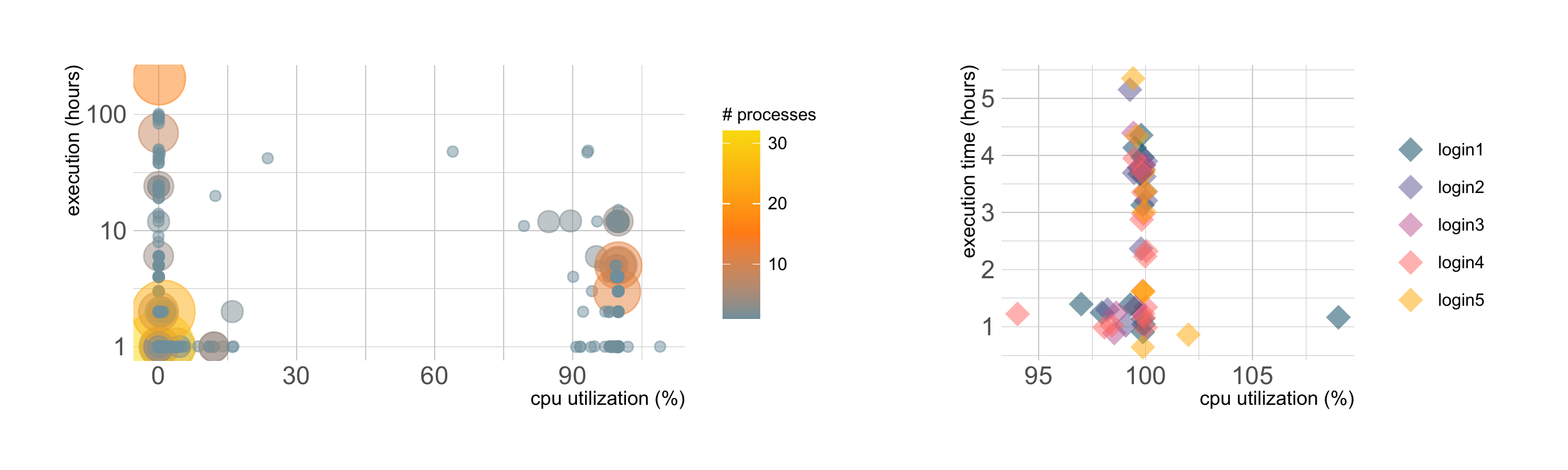}
    \vspace{-25pt}
    \caption{\emph{Left:} Execution of \texttt{mpirun} and \texttt{mpiexec} on
            login nodes (1,172 instances from 74 users).
        \emph{Right:} 56 executions of GROMACS (\texttt{gmx\_mpi}) on login
            nodes by two users. Each user runs instances of GROMACS on every
            login node, which occupy the available GPUs for several hours.}
    \label{fig:mpirun}
\end{figure}

\pp{Tightly Coupled Applications}
We have identified 1,172 uses of \texttt{mpirun} and
\texttt{mpiexec} by 74 users for running tightly coupled applications in the
login node. (Our filtering process removed mentions to compiling operations and
flags, environment variables, etc.) In further investigation, we
noticed that 816 out of the 1,172 instances of \texttt{mpirun} and
\texttt{mpiexec} spawned only a single process for less than one hour -- which
suggests that those executions were simple tests.
Figure~\ref{fig:mpirun}-\emph{left} shows execution times for the
\texttt{mpirun} and \texttt{mpiexec} instances, their associated CPU
utilization, and the number of processes spawned. The longest execution runs
for 204 hours and spawns 16 processes, followed by a dozen of executions that
run for about 100 hours. There is also a cluster of instances that consume more
than 90\% of CPU for an average of 12 hours, with two instances running for 47
and 49 hours each. A detailed look at these instances unveiled that
they use up to 4 cores from the login nodes and up to 4 GB of RAM each, which
could then considerably impact the performance of sound processes (compilation,
(de)compression, file synchronization, etc.) from other users.

Figure~\ref{fig:mpirun}-\emph{right} shows a subset of the executions shown in
Figure~\ref{fig:mpirun}-\emph{left}, which corresponds to executions of 
GROMACS~\cite{van2005gromacs}, a widely used molecular dynamics package, on 
GPUs in the login nodes. Specifically, we highlight a use case in which two 
users attempt to ``game the system'' by launching concurrent executions of 
the GPU-enabled version of GROMACS (\texttt{gmx\_mpi}), configured to spawn 
one CPU process and as many GPU processes as available in the system. 
To prevent such behaviors, Summit enforces limits on the login nodes to 
ensure resource availability by leveraging the Linux kernel feature 
\texttt{cgroups}: each user is limited to 16 hardware threads, 16 GB of memory,
and 1 GPU; and after 4 hours of CPU-time all login sessions are limited to 0.5
hardware threads; after 8 hours, the process is automatically killed. These
limits are reset as new login sessions are started. These two users consumed
50\% of all GPU resources across login nodes for about 84 consecutive hours,
however, through a synchronized process in which each of them re-initiated a
session periodically, so the limits would be reset. This behavior is not only
substantially harmful to other users by preventing a fair share of resources,
but also it conflicts with best practices of not running scientific
applications within login nodes.

\begin{figure}[!t]
    \centering
    \vspace{-10pt}
    \includegraphics[width=\linewidth]{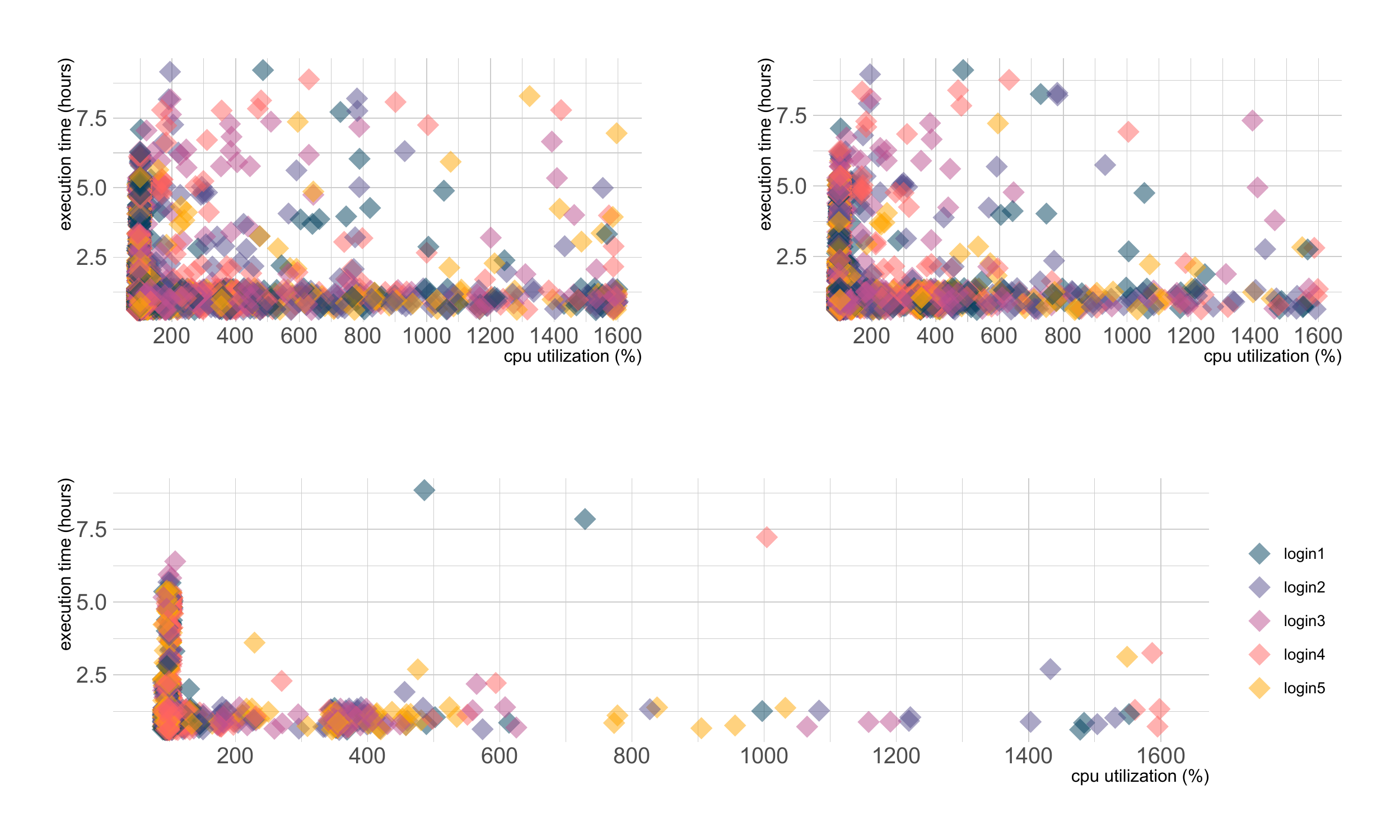}
    \vspace{-25pt}
    \caption{\emph{Left:} Execution of user processes (high-throughput
                applications) on login nodes. 
             \emph{Right:} Execution of Python programs on login nodes. 
             \emph{Bottom:} Execution of Python programs by users that have
                never submitted a batch job to the system.
            (Note that 1600\% CPU utilization means that a process comprising
            16 threads consumed 100\% CPU utilization each.)}
    \label{fig:htc}
\end{figure}

\pp{High-Throughput Applications}
We have identified a substantial number of executions of high-throughput 
applications on the login nodes. Here, we focus on a subset of these executions
that consumes more than 90\% of CPU per process, which comprises 8,014
instances executed by 549 users (27.9\% of total users).
Figure~\ref{fig:htc}-\emph{left} shows the distribution of user processes
\emph{vs.} their length, in hours, that run user codes (i.e., scientific
applications) on the login nodes. As for the tightly-coupled applications
above, we have filtered out all instances related to sound processes
(compilation, (de)compression, file synchronization, etc.). Users ran a wide
range of codes---495 unique programs---in which $\sim$78\% of them run for less
than an hour; thus, we consider them as execution tests. Some instances span 16
threads (\texttt{cgroups} limit) and run up to 7 hours, while others (about
7\% of the dataset) use more than 8 threads and run between 3 and 8 hours. We
then consider these instances as misuse of the login nodes. Due to the limits
imposed by Summit, we do not observe any attempt to ``game the system''; these
processes are mostly evenly distributed across login nodes, with a slightly
higher number for login3 (405 instead of 330 on average for the other login
nodes).

In spite of the large variation of user programs, we have identified that 4,478
instances (from 329 users) are running Python programs
(Figure~\ref{fig:htc}-\emph{right}). These instances represent 72.6\% of the
instances shown in Figure~\ref{fig:htc}-\emph{left}, which run for more than an
hour. This result indicates that some users may tend to use these login nodes
as additional computing resources, or even as their sole computing node. In
order to assert the latter, we attempted to isolate the list of users that ran
any of these codes without ever submitting a single job to the batch queue.
Astonishingly, we identified 41 users that fall into this category, which
comprises 1,012 instances, i.e., 12.6\% of the original dataset
(Figure~\ref{fig:htc}-\emph{bottom}). Although running user programs on login
nodes as an extension of computing resources is against best practices, using a
leadership-class HPC system for running user-based codes uniquely on login nodes
must be prevented---strict policies and processes should then be defined to
impede similar misuse of resources.

\medskip

While the \texttt{cgroups} mechanism protects the overall login node resources,
it falls short in ``low key'' and ``gaming the system'' misuse, as shown above.
Several measures may be taken to mitigate these issues. For example, the data
collected by this work can be used to identify misusers, either to educate them
about best practices or perhaps to introduce punitive actions. We will not
conjecture about potential new policies here, however.

\subsection{Scientific Workflows}

Scientific workflows are used almost universally across scientific domains for
solving complex and large-scale computing and data analysis problems. The
importance of workflows is highlighted by the fact that they have underpinned
some of the most significant discoveries of the past few
decades~\cite{badia2017workflows}. Many of these workflows have significant
demands for computation, storage, and communcation, and thus they have been
increasingly executed on large-scale computer
systems~\cite{ferreiradasilva-fgcs-2017}. In this section, we seek to
identify how and to what extent workflows have been used on Summit. Typically,
workflow systems run a coordinator process that manages workflow tasks'
dependencies, launches jobs to the batch queue as their dependencies are
satisfied, monitors their jobs' execution, and performs data movement
operations on behalf of the user. Table~\ref{tab:workflows} shows the total
number of processes run by workflow systems in Summit login nodes. In total, 71
users utilized workflow technologies for automating the execution of their
scientific applications. These processes often refer to agents that manage the
workflow execution and they can take several formats: from single orchestration
components (e.g., Swift/T) to the management of ensembles (e.g., RADICAL/EnTK).
The former leverages batch jobs for defining workflows within a parallel,
tightly coupled application (thus the lower number of processes), while the 
latter manages sets of tasks as high-throughput applications, i.e. the
so-called \emph{pilot jobs} \cite{pilots}.

\begin{table}[!t]
    \centering
    \setlength{\tabcolsep}{2.5pt}
    \scriptsize
    \caption{Total number of workflow management systems' processes observed
        across Summit login nodes (Jan 2020--Dec 2021).}
    \begin{tabular}{lrrrrrrrrrr}
        \toprule
        & parsl & swift/t & pegasus & fireworks & dask & maestro & cylc & dagman & snakemake & radical \\
        \midrule
        processes & 3,807 & 88 & 5,399 & 319 & 40,875 & 2,225 & 106 & 4 & 15,797 & 2,113,192 \\
        users & 7 & 3 & 3 & 5 & 27 & 6 & 1 & 1 & 5 & 13 \\
        \bottomrule
    \end{tabular}
    \label{tab:workflows}
\end{table}

\begin{figure}[!t]
    \centering
    \includegraphics[width=\linewidth]{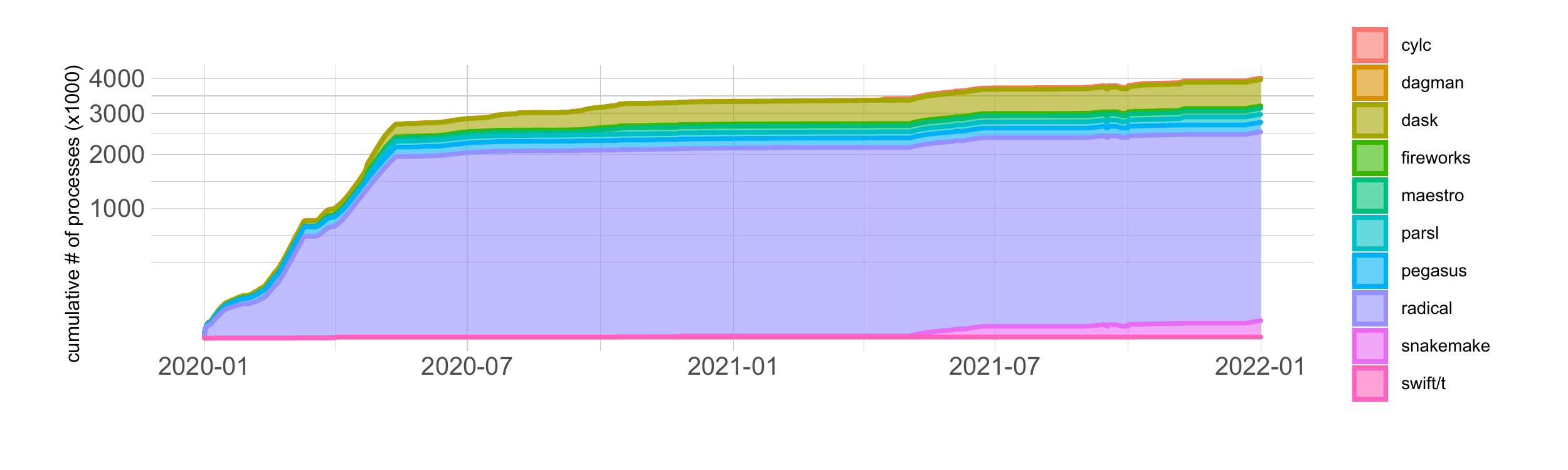}
    \vspace{-25pt}
    \caption{Cumulative number of workflow management systems' processes
            observed across Summit login nodes (Jan 2020--Dec 2021), shown with
            square root scale.}
    \label{fig:workflows}
\end{figure}

Figure~\ref{fig:workflows} shows the cumulative number of workflow-related 
processes across Summit login nodes for our dataset. Overall, workflow 
technology adoption has gradually increased throughout these past two years. 
A notable growth in workflow usage is observed in the first two quarters 
of 2020, which coincides with research conducted to understand the COVID-19 
pandemic through the use of HPC. Specifically, this research leveraged the 
RADICAL/EnTK framework for investigating spike dynamics in a variety of 
complex environments, including within a complete SARS-CoV-2 viral envelope 
simulation~\cite{casalino2021ai}. This research has been awarded the 2020 
ACM Gordon Bell Special Prize for High Performance Computing-Based COVID-19 
Research.

%% file: sec-relatedwork.tex
\section{Related Work}
\label{sec:relatedwork}

Analyzing and characterizing HPC workloads is a common practice for measuring
system and application performance metrics and thus identifying potential 
bottlenecks and atypical behaviors~\cite{feitelson2008looking}. For example,
the National Energy Research Scientific Computing Center (NERSC) has profiled 
and characterized three generations of their supercomputing
systems~\cite{nersc}. In these studies, HPC benchmarks are used to obtain
performance measurements, which are then used for the procurement process of
machines. Similarly, a characterization of the workload of Tianhe-1A at the
National Supercomputer Center in Tianjin presents equivalent system-level
metrics~\cite{feng2018workload}. In~\cite{10.1109/SC.2018.00077}, a
characterization of a parallel filesystem unveils I/O bottlenecks for different
classes of applications. Conversely, our analyses in this paper target users'
experience and behavior on login nodes---the interface to HPC systems.

In~\cite{schlagkamp-iccs-2016}, user behavior is studied with regards to think
time, the time between the completion of a job and the submission of the next 
job by the same user. Although this work leverages this same concept for 
defining user sessions, the study conducted in~\cite{schlagkamp-iccs-2016}
attempted to understand and characterize patterns of job submissions. Our
work, instead, seeks to understand user behavior on login nodes and 
relate their actions to misuses of the system or performance issues.
To the best of our knowledge, this is the first work that conducts such a
study.

%
%

%% file: sec-conclusion.tex
\section{Conclusion and Future Work}
\label{sec:conclusion}

We examined observation data from the login nodes of the leadership-class
Summit supercomputer at OLCF. We analyzed traditional system performance
metrics such as user access, I/O throughput, and job characteristics, as well
as user behavior regarding session lengths, misuse of login nodes, and how
users have leveraged workflows to perform complex, distributed computing. Our
findings identified key usage patterns that we believe will shed light on the
usage of login nodes on contemporary clusters and supercomputers. As immediate
future work, we will continue to collect this observation data for the rest of
the life of Summit, and we will start data collection for the upcoming exascale
Frontier supercomputer at OLCF. We also intend to analyze the data further into
other dimensions, including resource usage balancing and correlation of
external events (e.g., conference deadlines, call for proposals deadlines,
etc.), as well as the impact of the COVID-19 pandemic on the user behavior. 
